\documentclass[times,tighten,twocolumn]{aastex631}
\usepackage{amsmath}	
\usepackage{amssymb}	
\usepackage{acronym} 
\usepackage{longtable}
\usepackage{multirow}
\usepackage{units}
\usepackage{soul}

\accepted{\apj}

\shorttitle{Long-period Pulsars as Possible Outcomes of Supernova Fallback Accretion}
\shortauthors{Ronchi et al.}

\begin{document}

\title{Long-period Pulsars as Possible Outcomes of Supernova Fallback Accretion}



\correspondingauthor{Michele Ronchi}
 \email{ronchi@ice.csic.es}

\author[0000-0003-2781-9107]{M. Ronchi}
\affiliation{Institute of Space Sciences (ICE-CSIC), Campus UAB, Carrer de Can Magrans s/n, 08193, Barcelona, Spain}
\affiliation{Institut d'Estudis Espacials de Catalunya (IEEC), Carrer Gran Capit\`a 2--4, 08034, Barcelona, Spain}

\author[0000-0003-2177-6388]{N. Rea}
\affiliation{Institute of Space Sciences (ICE-CSIC), Campus UAB, Carrer de Can Magrans s/n, 08193, Barcelona, Spain}
\affiliation{Institut d'Estudis Espacials de Catalunya (IEEC), Carrer Gran Capit\`a 2--4, 08034, Barcelona, Spain}

\author[0000-0002-6558-1681]{V. Graber}
\affiliation{Institute of Space Sciences (ICE-CSIC), Campus UAB, Carrer de Can Magrans s/n, 08193, Barcelona, Spain}
\affiliation{Institut d'Estudis Espacials de Catalunya (IEEC), Carrer Gran Capit\`a 2--4, 08034, Barcelona, Spain}

\author[0000-0002-5119-4808]{N. Hurley-Walker}
\affiliation{International Centre for Radio Astronomy Research, Curtin University, Kent St, Bentley WA 6102, Australia}

\def\xmm {\emph{XMM--Newton}}
\def\cxo {\emph{Chandra}}
\def\nustar {\emph{NuSTAR}}
\def\rst {\emph{ROSAT}}
\def\swift {\emph{Swift}}
\def\nicer {\emph{NICER}}
\def\hxmt {\emph{Insight}-HXMT}
\def\pks {Parkes}

\def\flux {\mbox{erg\,cm$^{-2}$\,s$^{-1}$}}
\def\lum {\mbox{erg\,s$^{-1}$}}
\def\nh {N$_{\rm H}$}
\def\kms  {\rm \ km \, s^{-1}}
\def\cms  {\rm \ cm \, s^{-1}}
\def\gs   {\rm \ g  \, s^{-1}}
\def\cmtre {\rm \ cm^{-3}}
\def\cmdue {\rm \ cm$^{-2}$}
\def\ss {\mbox{s\,s$^{-1}$}}
\def\chisq {$\chi ^{2}$}
\def\rchisq {$\chi_{r} ^{2}$}

\def\arc{\mbox{$^{\prime\prime}$}}
\def\arcmin{\mbox{$^{\prime}$}}
\def\deg{\mbox{$^{\circ}$}}

\def\rsun {~R_{\odot}}
\def\msun {~M_{\odot}}
\def\mdotav {\langle \dot {M}\rangle }

\def\gleamfirst {\mbox{GLEAM-X\,J162759.5-523504.3}}
\def\gleam {\mbox{GLEAM-X\,J1627}}
\def\mtp{\mbox{PSR J0901-4046}}
\def\rcw{\mbox{1E\,161348-5055}}
\def\lowbsgr{\mbox{SGR\,0418+5729}}
\def\sgrfrb{SGR\,1935+2154}
\def\xte{XTE\,J1810$-$197}

\begin{abstract}
For about half a century the radio pulsar population was observed to spin in the $\sim$ 0.002--12~s range, with different pulsar classes having a spin-period evolution that differs substantially depending on their magnetic fields or past accretion history. The recent detection of several slowly rotating pulsars has re-opened the long-standing question of the exact physics, and observational biases, driving the upper bound of the period range of the pulsar population. In this work, we perform a parameter study of the spin-period evolution of pulsars interacting with supernova fallback matter and specifically look at the fallback accretion disk scenario. Depending on the initial conditions at formation, this evolution can differ substantially from the typical dipolar spin-down, resulting in pulsars that show spin periods longer than their coeval peers. By using general assumptions for the pulsar spin period and magnetic field at birth, initial fallback accretion rates and including magnetic field decay, we find that very long spin periods ($\gtrsim \unit[100]{s}$) can be reached in the presence of strong, magnetar-like magnetic fields ($\gtrsim \unit[10^{14}]{G}$) and moderate initial fallback accretion rates ($\sim \unit[10^{22-27}]{g \, s^{-1}}$). 
In addition, we study the cases of two recently discovered periodic radio sources, the pulsar \mtp\ ($P = \unit[75.9]{s}$) and the radio transient \gleamfirst\ ($P = \unit[1091]{s}$), in light of our model. 
We conclude that the supernova fallback scenario could represent a viable channel to produce a population of long-period isolated pulsars that only recent observation campaigns are starting to unveil. 

\end{abstract}

\section{Introduction} 
\label{sec:intro}

The spin-period distribution of the pulsar population reflects intrinsic properties of neutron star formation, early evolution, magnetic field decay, and age. Until a few years ago, the spin distribution of observed isolated pulsars was ranging between $\sim$ 0.002--12~s. At the fastest extreme, we have recycled millisecond pulsars (mostly in binaries), while the slowest extreme is populated by magnetars. The historical lack of isolated pulsars with periods $\gtrsim \unit[12]{s}$ has been intriguing and interpreted in different ways, ranging from the presence of a death line below which radio emission is quenched \citep{Ruderman1975, Bhattacharya1991, Chen1993}, to magnetic field decay coupled with the presence of a highly resistive layer in the inner crust \citep[possibly due to the existence of a nuclear pasta phase;][]{Pons2013}, as well as due to an observational bias caused by high band pass filters in radio searches (albeit not in X-ray searches). Although the main reason is uncertain, all of these effects likely contribute at some level to the observational paucity of long-period pulsars \citep{Wu2020}.

However, recent radio surveys, in particular thanks to new radio interferometers such as the LOw Frequency ARray \citep[LOFAR;][]{VanHaarlem2013}, MeerKAT \citep{Jonas2009}, Australian SKA Pathfinder \citep[ASKAP;][]{Hotan2021}, and the Murchison Widefield Array \citep[MWA;][]{Tingay2013,Wayth2018}, have started to uncover the existence of a new population of slowly rotating radio pulsars that challenge our understanding of the pulsar population and its evolution.

Two radio pulsars, PSR\,J1903$+$0433 \citep{Han2021} and PSR\,J0250$+$5854 \citep{Tan2018}, have been recently discovered with periods of $\unit[14]{s}$ and $\unit[23]{s}$, respectively. Moreover, a $\sim\unit[76]{s}$ radio pulsar with a magnetic field of $B \sim \unit[1.3\times10^{14}]{G}$ \citep[\mtp,][]{Caleb2022} and a peculiar radio transient with a periodicity of $\sim\unit[1091]{s}$  \citep[\gleamfirst;][]{Hurley-Walker2022a} have been discovered. This latter source is the only one uncertain in nature. In particular, it has a very variable flux, showing periods of ``radio outburst'' lasting a few months, a 90\% linear polarization, and a very spiky and variable pulse profile. Several interpretations have been advanced to explain the mysterious nature of this source. As argued in the discovery paper, its emission characteristics are typical of observed radio magnetars \citep{Kaspi2017, Esposito2020} (although its magnetic field is still poorly constrained). Alternatively, it could be a strongly magnetized white dwarf having spun down to the observed long spin period due to its larger moment of inertia \citep[see also][for a discussion on the nature of this source]{Tong2022}.
Furthermore, a few years ago the X-ray emitting neutron star \rcw\ at the center of the $\unit[2]{kyr}$ old supernova remnant (SNR) RCW103, with a measured modulation of $\sim \unit[6.67]{hr}$, showed a large magnetar-like X-ray outburst \citep{Rea2016, D'Ai2016}, demonstrating the source's isolated magnetar nature despite its long period and young age. 
 
In general studying the possible mechanisms that could lead to the formation of long-period pulsars is of great interest in order to understand their origin, eventual links with other types of neutron stars and possible connections to periodic activity of transient events such as fast radio bursts \citep[see for example][]{Beniamini2020, Xu2021}. One possible avenue that could lead to enhanced spin-down, especially in magnetars, is the presence of mass-loaded charged particle winds and outflows that could be particularly active after giant-flare episodes and temporarily expand the open magnetic flux region of the star \citep[see for example][]{Thompson2000, Tong2013, Beniamini2020}. 
Another possibility is the interaction of the neutron star with a fallback disk formed after the supernova explosion. 
Soon after their formation, neutron stars will necessarily witness fallback accretion with different accretion rates depending on the progenitor properties and explosion dynamics \citep{Ugliano2012, Perna2014, Janka2021}. If the fallback mass possesses sufficient angular momentum, it could form a long lasting accretion disk that will interact with the neutron star. For certain ranges of the initial spin period $P_0$, magnetic field $B_0$ and disk accretion rate $\dot{M}_{\rm d,0}$, fallback after the supernova explosion can substantially affect the pulsar spin evolution, in some cases slowing down the pulsar period significantly more than standard dipolar spin-down losses alone. 
In this context, the long period observed in \rcw\ has been interpreted as the spin period of a magnetar interacting with a fallback disk in a propeller state by numerous authors \citep{Li2007, Ho2017, Tong2016, Xu2019}. 
Moreover \citet{Chatterjee2000, Ertan2009, Benli2016} developed detailed numerical models of the interaction between a fallback disk and a neutron star to explain the emission of magnetars. 
Although appealing, this model struggles to explain several observed features of these types of sources, such as their burst and flaring activity \citep[see][for a review]{Mereghetti2013, Turolla2015, Kaspi2017, Esposito2021}. 
Overall, these studies on fallback disk accretion suggest that the presence or absence of a fallback disk around newly born neutron stars could be a determining factor for their evolution in the $P$-$\dot{P}$ diagram (where $P$ is the neutron star spin period and $\dot{P}$ is the spin period derivative with respect to time) as well as their emission properties and could further be relevant to explain the connections between different neutron star classes \citep[see for example][]{Alpar2001}.  

In this study we revisit the fallback scenario to specifically analyze the spin-period evolution of newly born pulsars that witness accretion from a fallback disk (\S\ref{sec:fallback_scenario}), and determine the parameter ranges that allow pulsars to experience efficient spin-down in the presence of magnetic field decay. Furthermore, we study the characteristics of the newly discovered long-period radio source \gleam\ (see \S\ref{sec:gleam}) and pulsar \mtp\ (see \S\ref{sec:mtp0013}) in the context of the fallback scenario in order to constrain their nature and evolution (see \S\ref{sec:discussion}). We provide a summary in \S\ref{summary} \footnote{Jupyter Notebooks to reproduce the plots and results of our paper are publicly available at \url{https://github.com/MicheleRonchi/pulsar_fallback}}.

\section{Period evolution of pulsars slowing down via dipolar losses} 
\label{sec:dipolar_losses}

\begin{figure}
\centering
\includegraphics[width =0.48\textwidth]{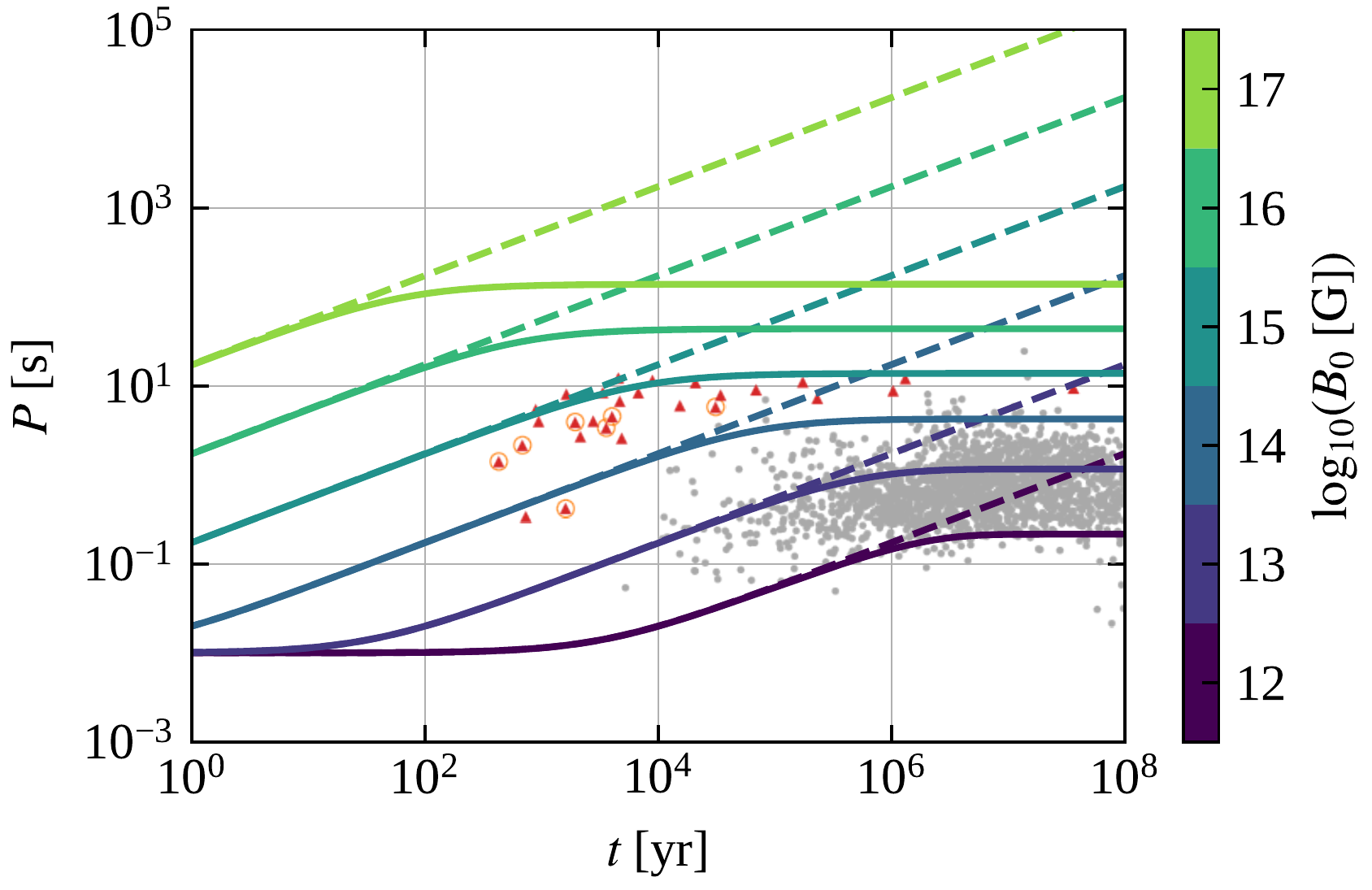}
\caption{Pulsar spin-period evolution in time for dipolar spin-down for different values of the initial magnetic field $B_0$. The dashed lines represent the evolution curves for a constant magnetic field, while the solid lines represent the evolution tracks for a decaying magnetic field according to eq. \ref{eq:B_decay}. In the background, the gray points represent the observed radio pulsars \citep[Data from the ATNF Pulsar Catalog, \url{https://www.atnf.csiro.au/research/pulsar/psrcat/};][]{Manchester2005} and red triangles represent the currently detected magnetars \citep[Data from the Magnetar Outburst Online Catalogue, \url{http://magnetars.ice.csic.es/};][]{CotiZelati2018} highlighting those that exhibit radio emission (orange circles). We consider the characteristic age $\tau_{\rm c} = P/(2\dot{P})$ as a proxy for their real age}.
\label{fig:em_spin_down_P_evol} 
\end{figure}  

Ordinary rotation powered pulsars are expected to slow down via electromagnetic dipolar losses, with an additional component driven by their magnetic field decay.
In this scenario, if $\omega = 2 \pi / P$ is the angular spin frequency of the pulsar, the electromagnetic torque causes the star to slow down according to: 

\begin{align}
    \frac{d\omega}{dt} = - \beta B(t)^2 \omega^3,
        \label{eq:domegadt}
\end{align}

\noindent where $c$ is the speed of light and for an aligned rotator $\beta \sim R_{\rm NS}^6/(4 c^3 I_{\rm NS}) \sim  \unit[1.5 \times 10^{-41}]{s \, G^{-2}}$ assuming a typical neutron star radius $R_{\rm NS} \sim \unit[11]{km}$, and moment of inertia $I_{\rm NS} \sim \unit[1.5 \times 10^{45}]{g \, cm^2}$. The inclination angle dependence and uncertainty on the neutron star mass and radius introduce a correction of at most an order of magnitude to the value of $\beta$ in the equation above \citep[see for example][]{Spitkovsky2006}.

If we consider a constant magnetic field $B(t) = B_0$, we can solve eq. \eqref{eq:domegadt} to find the spin-period evolution in time:

\begin{align}
    P(t) = P_0 \left(1 + \frac{t}{t_{\rm em}} \right)^{1/2}.
        \label{eq:P_t_em_spindown}
\end{align}
where we define an electromagnetic spin-down timescale $t_{\rm em} = P_0^2 / \left( 8 \pi^2 \beta B_0^2 \right)$.

However, several studies have shown that neutron star magnetic fields evolve over time and decay due to the combined action of Ohmic dissipation and the Hall effect in the star's crust \citep{Pons2007, Pons2009, Vigano2013, Pons2019, DeGrandis2020}. For simplicity, if we consider a crustal-confined magnetic field, we can use the phenomenological description of the field decay presented in \citet{Aguilera2008a, Aguilera2008b}, given by the analytic expression: 
\begin{align}
    B(t) = B_0 \frac{e^{-t/\tau_{\rm Ohm}}}{1 + \frac{\tau_{\rm Ohm}}{\tau_{\rm Hall,0}} \left[ 1 - e^{-t/\tau_{\rm Ohm}} \right]}. 
        \label{eq:B_decay}
\end{align}
This equation captures a first stage that is characterized by rapid (non-exponential) decay and regulated by the Hall timescale $\tau_{\rm Hall,0}$, and a second stage that is characterized by exponential decay due to Ohmic dissipation and regulated by the timescale $\tau_{\rm Ohm}$.
These two characteristic timescales can be defined by the following expressions:
\begin{align} \nonumber
    \tau_{\rm Hall,0} &= \frac{4 \pi e n_e L^2}{c B_0} \\ 
    &\simeq \unit[6.4 \times 10^4]{yr} \left( \frac{n_e}{\unit[10^{35}]{cm^{-3}}} \right)
    \left( \frac{L}{\unit[1]{km}} \right)^2 
    \left( \frac{B_0}{\unit[10^{14}]{G}} \right)^{-1}, 
        \label{eq:tau_hall} \\
    \tau_{\rm Ohm} &= \frac{4 \pi \sigma L^2}{c^2} \nonumber \\ 
    &\simeq \unit[4.4 \times 10^6]{yr} \left( \frac{\sigma}{\unit[10^{24}]{s^{-1}}} \right)
    \left( \frac{L}{\unit[1]{km}} \right)^2, 
        \label{eq:tau_ohm}
\end{align}
where $e$ is the electron charge, $n_e$ is an average electron density in the neutron star crust, $\sigma$ is the dominant conductivity based on phonon or impurity scattering and $L$ is the typical lengthscale over which the relevant physical quantities (i.e., $n_e$, $B_0$ and $\sigma$) change inside the crust \citep[see][]{Cumming2004, Gourgouliatos2014a}. 
Note also that in eq. \eqref{eq:B_decay}, $\tau_{\rm Hall,0}$ represents the Hall timescale for the initial magnetic field strength $B_0$ \citep[we refer to][for more details]{Aguilera2008b}. 
In general the value of the Hall and the Ohmic timescales can vary by orders of magnitude within the crust and during the evolution, depending strongly on the density profile and magnetic field intensity and curvature. The Ohmic timescale additionally varies with temperature due to its dependence on the conductivity. 
Although a rough simplification, eq. \eqref{eq:B_decay} and the timescales defined above are consistent with the evolution history inferred for Galactic magnetars \citep{Colpi2000, Dall'Osso2012, Beniamini2019} and are able to qualitatively capture the magnetic field evolution obtained with more sophisticated magneto-thermal numerical simulations \citep[see for example][]{Vigano2013, DeGrandis2020, Vigano2021}.
Using these prescriptions for the magnetic field, we solve eq. \eqref{eq:domegadt} and determine the evolution of the spin-period in time. In Fig. \ref{fig:em_spin_down_P_evol}, we present the corresponding behaviour for $B_0$ values in the range $\unit[10^{12-17}]{G}$. Since the long-term evolution for $t \gg t_{\rm em}$ is insensitive to the initial spin period value, we fix $P_0$ to a fiducial value of $\unit[10]{ms}$ which is compatible with the birth spin-period distributions inferred from population synthesis studies \citep[see for example][]{Faucher2006, Gullon2014, Cieslar2020}. After an initial phase of duration $t_{\rm em}$, where the spin period remains almost constant and equal to its initial value, $P$ starts to evolve $\propto t^{1/2}$. If the magnetic field remains constant, this evolution proceeds indefinitely (dashed lines). If $B$ decays over time according to eq. \ref{eq:B_decay}, the electromagnetic torque eventually becomes negligible and the spin period stops increasing and stabilizes (solid lines).

From this plot it is evident that, if one assumes a decaying magnetic field in the neutron star crust, explaining the existence of slowly rotating neutron stars with spin periods $P \gtrsim \unit[100]{s}$ becomes problematic since one would require rather extreme magnetic field values ($B \gtrsim \unit[10^{16}]{G}$). If instead some mechanism prevents the crustal magnetic field from decaying in time, the star could reach longer spin periods more easily, since a strong electromagnetic torque is maintained over longer times. After a few Hall timescales, the magnetic field in the crust reorganizes towards smaller scales through the Hall cascade \citep[e.g.][]{Brandenburg2020} and approaches a quasi-equilibrium configuration during which the dissipation of the magnetic field is slowed down. In this phase, commonly named the ``Hall attractor'' \citep{Gourgouliatos2014b}, the electric currents are predominantly confined to the inner crust where subsequent dissipation occurs on the Ohmic timescale, which in turn depends on the local properties of the inner crust. As a result, magnetic field decay cannot be halted indefinitely but will proceed on timescales of a few Myr \citep[see however][on how the pasta phase could greatly shorten these timescales]{Pons2013}.
Another possibility for maintaining strong magnetic fields is that the electric currents could be predominantly present in the neutron star core allowing the magnetic field to stay stable and barely decay over time \citep[see for example][]{Vigano2021}. However it is still unclear if such conditions can be realized since little is known about magnetic field evolution in neutron star cores \citep[in particular the effect of superfluid and superconducting components on field evolution;][]{Graber2015, Passamonti2017, Ofengeim2018, Gusakov2020}. In general we cannot exclude the possibility that long-period pulsars are the result only of electromagnetic spin-down in the presence of strong and persistent magnetic fields, possibly supported by a Hall attractor or a core component. However, in light of our current knowledge of magnetic field evolution in neutron stars, we suggest that dipolar spin-down alone struggle to explain the existence of long-period pulsars. Instead, we argue below that these sources could be the result of a different scenario, whose ingredients are readily available in standard neutron-star formation models.   

\section{Period evolution of pulsars witnessing supernova fallback} 
\label{sec:fallback_scenario}

An alternative scenario that could explain the existence of strongly magnetized and slowly rotating pulsars involves the interaction between a newly born highly magnetic neutron star and fallback material from the supernova explosion. In the following for simplicity, we assume that the star's magnetic axis is aligned with the rotation axis and that the magnetic field is crust-dominated and decays in time according to eq. \eqref{eq:B_decay}.
In the early stages after the supernova explosion, the proto-neutron star emits a powerful neutrino wind which exerts a pressure on the outer envelope of the progenitor star. The duration of this wind is believed to be of the order of $\sim \unit[10]{s}$ after core bounce, which corresponds to the neutrino-cooling timescale for a newborn neutron star \citep{Ugliano2012, Ertl2016}. On this timescale, the magnetosphere of the neutron star reaches an equilibrium configuration in the region swept by the neutrino wind. After the subsiding of this neutrino-driven wind, the gas in the inner envelopes of the exploding star decelerates due to the gravitational pull of the neutron star and collapses back. This leads to the onset of fallback accretion. According to simulations \citep{Ugliano2012, Ertl2016, Janka2021}, the total fallback mass can reach values up to $M_{\rm fb} \lesssim 0.1 M_{\odot}$, while the fallback mass rate can reach values of around $\sim \unit[10^{27-31}]{g \, s^{-1}}$ in the first $\sim \unit[10-100]{s}$ after bounce, afterwards decreasing according to a power law $\sim t^{-5/3}$, which is compatible with theoretical predictions for spherical supernova fallback \citep{Michel1988, Chevalier1989}. However, if part of the fallback matter possesses sufficient angular momentum, it will circularize to form an accretion disk. Hereafter, we will study in detail the case where a disk forms successfully and interacts with the central neutron star. We assume that the accretion disk's rotation is prograde with respect to the spin of the neutron star and consider, for simplicity, a magnetic dipole moment aligned with the neutron star rotation axis. We note that in order to observe pulsed radio emission from sources of this kind, a misalignment between the magnetic axis and the spin axis is, in principle, required. We do not study the impact of this effect in our calculations but point out that for small misalignment angles the evolutionary scenario should not differ substantially from the one studied in this work. At the end of this section, we briefly discuss the case where the conditions to form a disk are not met and fallback is expected to proceed almost spherically.

\subsection{Accretion from a fallback disk}

\citet{Mineshige1997} and \citet{Menou2001} have studied the formation and time evolution of fallback accretion disks around compact objects. In particular, by using the self-similar solution of \citet{Cannizzo1990}, \citet{Menou2001} found that the fallback material with excess angular momentum circularizes to form a disk on a typical viscous timescale $t_{\rm v} \sim \unit[2.08 \times 10^3 \, T_{\rm c, 6}^{-1} r_{\rm d, 8}^{1/2}]{s} $, where $T_{\rm c, 6}$ is the disk's central temperature in units of $\unit[10^6]{K}$ and $r_{\rm d, 8}$ is the circularization radius of the disk in units of $\unit[10^8]{cm}$. The circularization radius $r_{\rm d}$ can be assumed to be the Keplerian radius corresponding to the initial angular momentum of the fallback matter. Supernova simulations have shown that typical values for the angular momentum density of the fallback matter are around $j_{\rm fb} \sim \unit[10^{16-17}]{cm^2 \, s^{-1}}$ \citep{Janka2021}, which corresponds to a circularization radius of around $\unit[10^{6-8}]{cm}$. In the following we will always assume the fiducial values $T_{\rm c, 6} = 1$ and $r_{\rm d, 8} = 1$ \citep[see also][]{Hameury1998}. The viscous timescale determines the duration of an initial transient accretion phase characterized by a nearly constant accretion rate. Afterwards, as the supply of fallback matter decreases, the accretion rate into the disk itself declines as a power law. Furthermore the disk starts to spread due to viscous effects. We follow these prescriptions and model the long-term time evolution of the accretion rate and outer radius of the disk as \citep[see][]{Menou2001, Ertan2009}:
\begin{align} \label{eq:disk_rate_evolution}
    \dot{M}_{\rm d}(t) = \dot{M}_{\rm d, 0} \left( 1 + \frac{t}{t_{\rm v}} \right)^{-\alpha},
\end{align}
\begin{align} \label{eq:outer_radius_evolution}
    r_{\rm out}(t) = r_{\rm d} \left( 1 + \frac{t}{t_{\rm v}} \right)^{\gamma},
\end{align}
\noindent where the coefficients $\alpha$ and $\gamma$ depend on the main mechanism determining the opacity of the disk. In particular $\alpha = 19/16 \simeq 1.18$ and $\gamma = 3/8 \simeq 0.38$ if the disk opacity is dominated by electron scattering or $\alpha = 5/4 = 1.25$ and $\gamma = 1/2$ if the disk is dominated by Kramer's opacity \citep{Cannizzo1990, Menou2001, Ertan2009}. In the following, we adopt intermediate values of $\alpha = 1.2$ and $\gamma = 0.44$. In general, determining the fraction of fallback matter that eventually forms the disk is not trivial, since this depends on the progenitor's properties and on the supernova mechanism itself. Here, we consider a broad range of values for the disk accretion rate between $\unit[10^{19-29}]{g \, s^{-1}}$ that is compatible with the disk accreting at a fraction of the overall supernova fallback rate as well as having a fraction of the total fallback mass $M_{\rm fb}$. For a radiatively efficient disk, if this flow of matter is maintained constant throughout the disk up to the inner disk radius $r_{\rm in}$, such accretion rates would result in luminosities up to $L_{\rm acc} \simeq G M_{\rm NS} \dot{M}_{\rm d}/(2 r_{\rm in}) \simeq \unit[10^{46}]{erg \, s^{-1}}$ for fiducial values $M_{\rm NS} = 1.4 M_{\odot}$, $\dot{M}_{\rm d} = \unit[10^{28}]{g \, s^{-1}}$ and $r_{\rm in} = \unit[10^8]{cm}$; $G$ is the gravitational constant. These luminosities far exceed the Eddington limit given by $L_{\rm Edd} = \unit[1.3 \times 10^{38} (M_{\rm NS} / M_{\odot})]{erg \, s^{-1}} \simeq \unit[1.8 \times 10^{38}]{erg \, s^{-1}}$. It is thus expected that such super-critical inflows of matter produce winds and outflows that reduce the accretion rate in the inner disk region to values below the Eddington limit \citep{poutanen2007}. 
We note that in the presence of strong magnetic fields ($B \gtrsim \unit[10^{14}]{G}$) the Eddington limit could be increased due to a reduction in the electron-scattering opacity \citep{Canuto1971, Bachetti2014}. In our model, we neglect this effect as it is relevant only in the very early stages of the evolution when the inner disk radius closely approaches the neutron star surface but it does not significantly affect the long-term dynamics outlined below.
In a simplified scenario, we hence assume that the accretion rate at the inner disk radius $r_{\rm in}$ has to be limited by the Eddington accretion rate, given by $\dot{M}_{\rm Edd} \simeq 2 L_{\rm Edd} r_{\rm in} / (G M_{\rm NS}) \simeq \unit[10^{18} (r_{\rm in}/R_{\rm NS})]{g \, s^{-1}}$. 
Therefore, we model the accretion rate at $r_{\rm in}$ as:

\begin{align} \label{eq:accretion_rate_evolution}
    \dot{M}_{\rm d, in} = 
        \begin{cases}
          \dot{M}_{\rm Edd} & \text{if $\dot{M}_{\rm d} \geq \dot{M}_{\rm Edd}$}, \\
          \dot{M}_{\rm d}   & \text{if $\dot{M}_{\rm d} < \dot{M}_{\rm Edd}$}.
        \end{cases}
\end{align}

If this accretion rate is sufficiently high, the in-falling matter is able to deform and penetrate the closed magnetosphere whose boundary can be roughly defined by the light cylinder radius $r_{\rm lc}$:

\begin{align} \label{eq:light_cylinder_radius}
    r_{\rm lc} = \frac{c}{\omega}.
\end{align}

\noindent In this case, we define the magnetospheric radius $r_{\rm m}$ which represents the distance from the star where the magnetic pressure equals the ram pressure of the accreted flow \citep{Davidson1973, Elsner1977, Ghosh1979}:

\begin{align} \label{eq:magnetospheric_radius}
    r_{\rm m} = \xi \left( \frac{\mu^4}{2 G M_{\rm NS} \dot{M}_{\rm d, in}^2} \right)^{\frac{1}{7}} ,
\end{align}

\noindent where $\xi \sim 0.5$ is a corrective factor that takes into account that the accretion disk has a non-spherical geometry \citep{Long2005, Bessolaz2008, Zanni2013}, $\mu = B(t) R_{\rm NS}^3 / 2$ is the stellar magnetic moment, and $B(t)$ is the time-dependent magnetic field at the pole (assuming a dipolar structure for the magnetosphere). Note that the position of the magnetospheric radius evolves in time due to the decay of both the disk accretion rate and of the magnetic field. Inside the magnetospheric radius, the plasma dynamics are dominated by the magnetic field so that the accretion flow is forced to corotate with the star's closed magnetosphere. Under such conditions, the magnetospheric radius roughly determines the inner edge of the accretion disk so that we can assume $r_{\rm in} \sim r_{\rm m}$. 

If the accretion rate is sufficiently low, the magnetic pressure of the closed magnetosphere is able to keep the accretion flow outside the light cylinder. However beyond $r_{\rm lc}$, the dipole configuration breaks down, the magnetic field lines open up and as a consequence they are no longer able to exert significant pressure on the accreted plasma. Under these conditions, we expect the inner radius $r_{\rm in}$ of the disk to be roughly equal to $r_{\rm lc}$. Thus, in general we adopt the same prescription as \citet{Yan2012} and assume $r_{\rm in} \simeq \min(r_{\rm m}, r_{\rm lc})$.
Note that a condition for the disk to form and remain active is that the disk's outer radius satisfies $r_{\rm out} > r_{\rm in}$. Otherwise, the stellar magnetic field would completely disrupt the disk, restoring a configuration where the neutron star's spin-down is determined by dipolar losses only.

Another critical lengthscale is the corotation radius $r_{\rm cor}$, which represents the distance at which the gravitational pull of the neutron star balances the centrifugal force for a test mass that is corotating with the star at the spin frequency $\omega$:

\begin{align} \label{eq:corotation_radius}
    r_{\rm cor} = \left( \frac{G M_{\rm NS}}{\omega^2} \right)^{\frac{1}{3}}.
\end{align}

The position of the magnetospheric radius with respect to the other two radii determines the total torque $N_{\rm tot}$ exerted on the star, which in turn drives the time evolution of the spin period according to:

\begin{align} \label{eq:omega_evolution}
    I_{\rm NS} \dot{\omega} = N_{\rm tot} = N_{\rm acc} + N_{\rm dip},
\end{align}

\noindent where $N_{\rm dip}$ accounts for the electromagnetic torque of the magnetosphere and $N_{\rm acc}$ accounts for the torque exerted by the accretion process. Following \citet{Piro2011, Metzger2018} the total torque can be modeled by the following equation:

\begin{align} \label{eq:torque}
    N_{\rm tot} = 
        \begin{cases}
          \dot{M}_{\rm d, in} r_{\rm lc}^2 \left[ \Omega_{\rm K}(r_{\rm lc}) - \omega \right] \\ \quad
            - I_{\rm NS} \beta B^2 \omega^3   & \text{if $r_{\rm m} > r_{\rm lc}$}, \\
          \dot{M}_{\rm d, in} r_{\rm m}^2 \left[ \Omega_{\rm K}(r_{\rm m}) - \omega \right] \\ \quad
            - I_{\rm NS} \left( \frac{r_{\rm lc}}{r_{\rm m}} \right)^2 \beta B^2 \omega^3 & \text{if $r_{\rm m} \leq r_{\rm lc}$}.
        \end{cases}
\end{align}
where the first term is the accretion torque $N_{\rm acc}$ and the second term is the dipolar torque $N_{\rm dip}$.

\begin{figure}
\centering
\includegraphics[width = 0.48\textwidth]{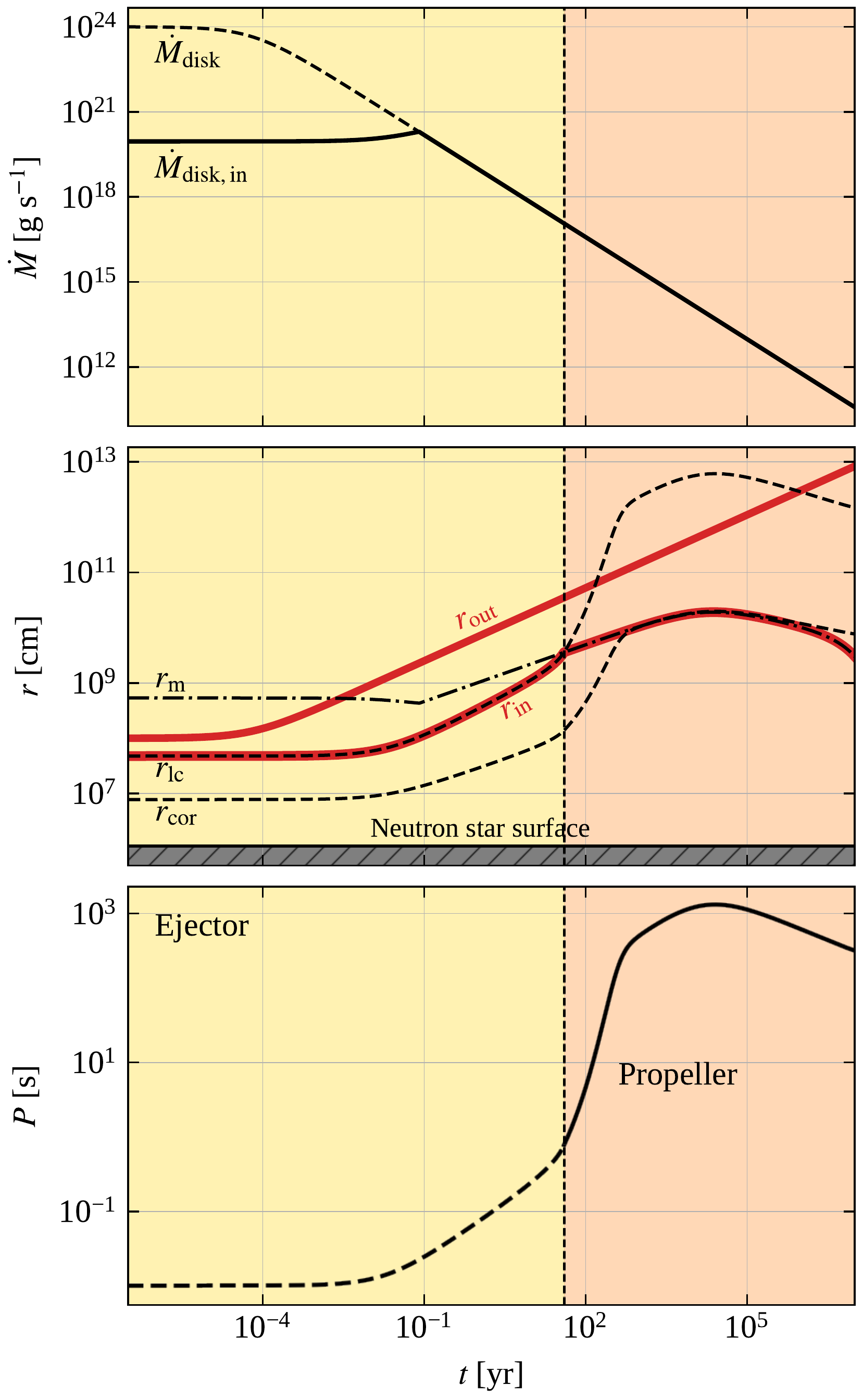}
\caption{Example of neutron star spin-down in presence of a fallback disk on a timescale of $\unit[10^7]{yr}$. The top panel shows the total disk accretion $\dot{M}_{\rm d}$ and the Eddington-limited accretion rate at the inner radius $\dot{M}_{\rm d, in}$. The middle panel illustrates the evolution of the three critical radii $r_{\rm cor}$, $r_{\rm m}$, $r_{\rm lc}$. The evolution of the disk's inner and outer radii is highlighted by the red lines. The bottom panel shows the resulting time evolution of the spin period. We assume an initial spin period $P_0 = \unit[10]{ms}$, initial magnetic field $B_0 = \unit[4 \times 10^{14}]{G}$ and an initial disk accretion rate $\dot{M}_{\rm d,0} = \unit[10^{24}]{g \, s^{-1}}$. We also highlight the duration of the ejector and propeller phases by shading the background in yellow and orange, respectively.} 
\label{fig:prop_evolution_ex}
\end{figure}  

Fig. \ref{fig:prop_evolution_ex} shows an example of a solution of eq. \eqref{eq:omega_evolution} for a pulsar interacting with a fallback disk. We consider an initial spin period $P_0 = \unit[10]{ms}$, initial magnetic field $B_0 = \unit[4 \times 10^{14}]{G}$ and an initial disk accretion rate $\dot{M}_{\rm d,0} = \unit[10^{24}]{g \, s^{-1}}$. Note that for this choice of initial parameters the disk can form and stays active since $r_{\rm out} > r_{\rm in}$ at all times. 
Furthermore in the early phases, accretion at the inner radius of the disk is limited by the Eddington limit. 

Different regimes are present depending on the relative ordering of the three radii defined above. In particular with reference to Fig. \ref{fig:prop_evolution_ex}, we distinguish:

\begin{itemize} 

\item For $r_{\rm m} > r_{\rm lc} > r_{\rm cor}$, the accreted material remains at the boundary of the closed magnetosphere and does not influence the neutron star's internal dynamics, i.e., the star spins down mainly due to dipolar electromagnetic torques, that is $N_{\rm tot} \sim N_{\rm dip}$. After a phase of duration $t_{\rm em}$ where the spin period stays constant, $P$ starts to increase $\propto t^{1/2}$ (see eq. \eqref{eq:P_t_em_spindown}). As a consequence the characteristic radii $r_{\rm lc}$ and $r_{\rm cor}$ increase as $t^{1/2}$ and $t^{2/3}$, respectively. This phase is commonly referred to as the ejector phase (shaded yellow region in Fig. \ref{fig:prop_evolution_ex}). The mechanism responsible for radio emission can be active and the neutron star could be observed as a radio pulsar. However, as $\dot{M}_{\rm d}$ decreases and eventually becomes sub-Eddington, the magnetospheric radius grows slower than the other two critical radii ($\propto t^{2 \alpha/7} \simeq t^{0.34}$ for $\alpha=1.2$). Thus, $r_{\rm m}$ will eventually cross the light cylinder.

\item For $r_{\rm lc} > r_{\rm m} > r_{\rm cor}$, the accretion flow is able to penetrate inside the closed magnetosphere. As it reaches the magnetospheric radius, the plasma flow is forced to corotate with the magnetosphere at super-Keplerian speeds causing it to be ejected due to centrifugal forces. This introduces a viscous torque that spins down the star very efficiently; a phase commonly referred to as the propeller (shaded orange region in Fig. \ref{fig:prop_evolution_ex}). In this case, we have $\omega > \Omega_K(r_{\rm m})$, where $\Omega_{\rm K}(r) = \left( G M_{\rm NS} / r^3 \right)^{1/2}$ is the Keplerian orbital angular velocity at radius $r$.
The electromagnetic torque $N_{\rm dip}$, i.e., the second term in eq. \eqref{eq:torque}, is also influenced by accretion. When the accretion flow penetrates inside the closed magnetosphere, the fraction of the magnetic field lines, which connects the star to the disk, is forced to open and as a consequence the polar cap region containing open field lines expands. This enhances the spin-down torque caused by the magnetosphere \citep{Parfrey2016, Metzger2018}. In particular \citet{Parfrey2016} argue that for $r_{\rm m} < r_{\rm lc}$, the magnetic flux through the expanded polar cap increases by a factor $\sim (r_{\rm lc}/ r_{\rm m})$. As a consequence, since the dipole spin-down torque is proportional to the square of the magnetic flux through the open field-line region, $N_{\rm dip}$ is enhanced by a factor $\sim (r_{\rm lc}/ r_{\rm m})^2$.
Moreover, since the radio emission is believed to be associated with magnetospheric currents and pair production \citep{Beloborodov2008, Philippov2020}, as the closed-magnetosphere geometry is disturbed by the accretion flow, the mechanism generating the radio emission is likely perturbed or even stopped, effectively switching off the radio-loud nature of these sources \citep{Li2006}. Therefore, we expect a neutron star in the propeller regime unlikely to be observable as a radio pulsar.
During this propeller phase, the spin frequency decreases over time and eventually $\omega \sim \Omega_K(r_{\rm m})$ (i.e., when $r_{\rm cor} \sim r_{\rm m}$), so that the net torque exerted on the neutron star vanishes and a spin equilibrium is reached. From this point onward, further evolution of the spin period will be regulated mainly by time variation of the accretion rate in the inner disk $\dot{M}_{\rm d, in}$ and the stellar magnetic field $B$. For example, the decay of the magnetic field strength causes the slow decrease of the magnetospheric radius at late times. This drives a progressive shift of the spin-equilibrium radius towards the neutron star surface, causing the neutron star to slightly spin up; in Fig. \ref{fig:prop_evolution_ex}, this happens when the magnetic field starts to decay after a Hall timescale, i.e., after $\sim \unit[10^4]{yr}$.

\item For $r_{\rm lc} > r_{\rm cor} > r_{\rm m}$, we have $\omega < \Omega_K(r_{\rm m})$. The accretion flow still manages to penetrate inside the closed magnetosphere but as it reaches the magnetospheric radius, the plasma is forced to corotate with the magnetosphere at sub-Keplerian speeds. In other words, at the boundary defined by the magnetospheric radius, the accreted plasma is orbiting faster than the neutron star magnetosphere. This introduces a viscous torque that transfers angular momentum from the plasma flow to the star and tends to spin up the star. Besides this, the enhanced electromagnetic spin-down torque described above is still acting, opposing the spin up due to accretion. Therefore, the evolution in this regime is controled by the relative strength of $N_{\rm acc}$ and $N_{\rm dip}$. In the example in Fig. \ref{fig:prop_evolution_ex}, this phase is experienced at very late times, i.e. $\sim \unit[10^6]{yr}$, when the magnetic field has decayed so much that the star exits the spin equilibrium and the magnetospheric radius becomes smaller than the corotation radius.

\end{itemize}

From these considerations, we observe that the parameters that mainly regulate the evolution of a pulsar surrounded by a fallback disk are the initial stellar magnetic field $B_0$ and the initial disk accretion rate $\dot{M}_{\rm d,0}$. The latter determines the time at which the accretion rate affecting the compact object becomes sub-Eddington and starts decreasing. In contrast, the value of the neutron star's initial spin period only affects the early evolution stages, while the long-term evolution of the neutron star is almost insensitive to the value of $P_0$. 

\begin{figure*}
\centering
\includegraphics[width = 1\textwidth]{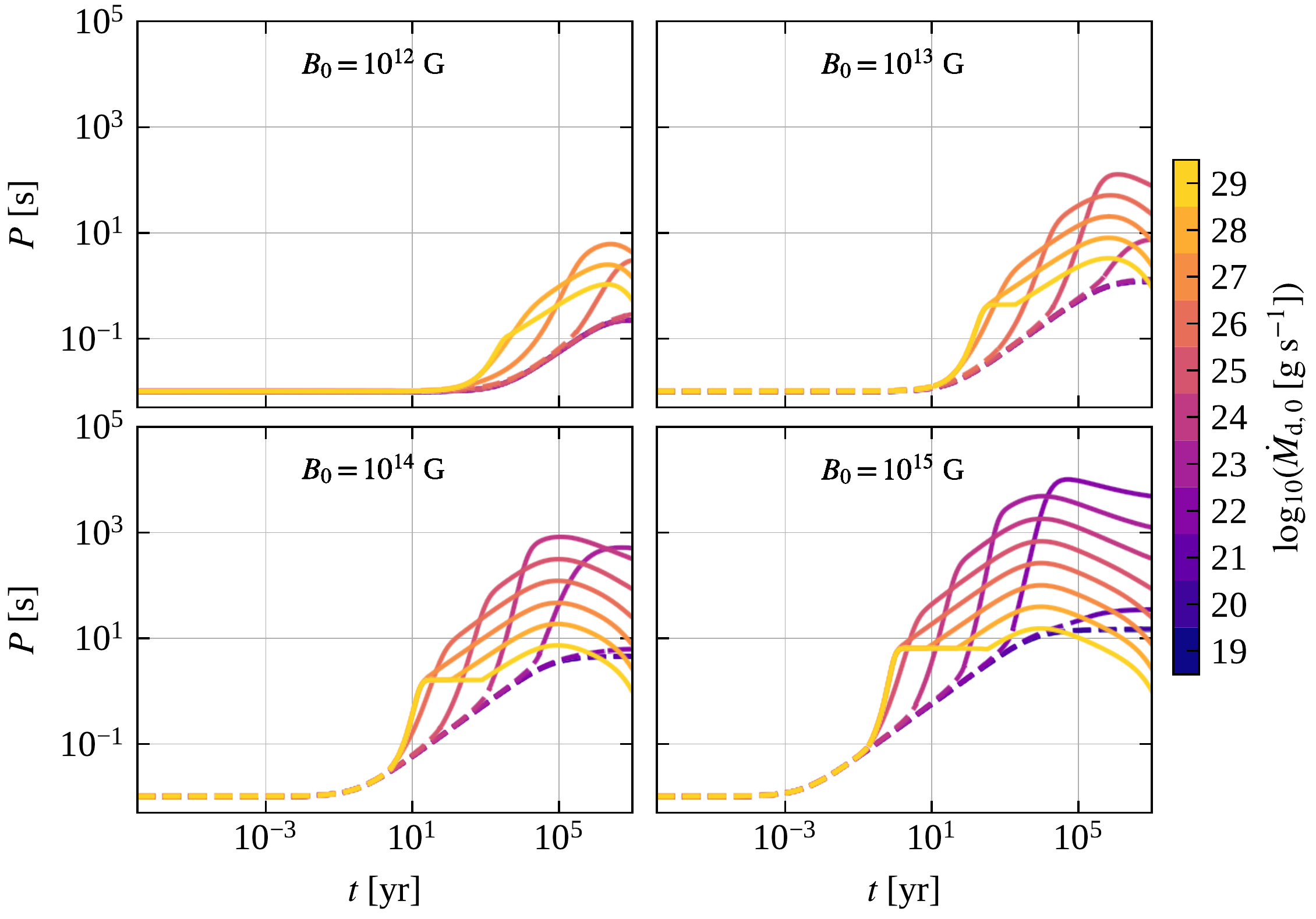}
\caption{Example curves showing the time evolution of the spin period for a pulsar interacting with a fallback disk on a timescale of $\unit[10^7]{yr}$, for different assumptions on the initial $B_0$ field strength, and varying disk fallback rate $\dot{M}_{\rm d,0}$. The dashed portion of the curves indicates when the neutron star is in the radio-loud ejector phase, i.e., when $r_{\rm m} > r_{\rm lc}$, while the solid portion indicates when the neutron star is in the radio-quiet propeller stage, i.e., when $r_{\rm m} < r_{\rm lc}$.}
\label{fig:P_evolution_B}
\end{figure*}  

\begin{figure*}
\centering
\includegraphics[width = 1\textwidth]{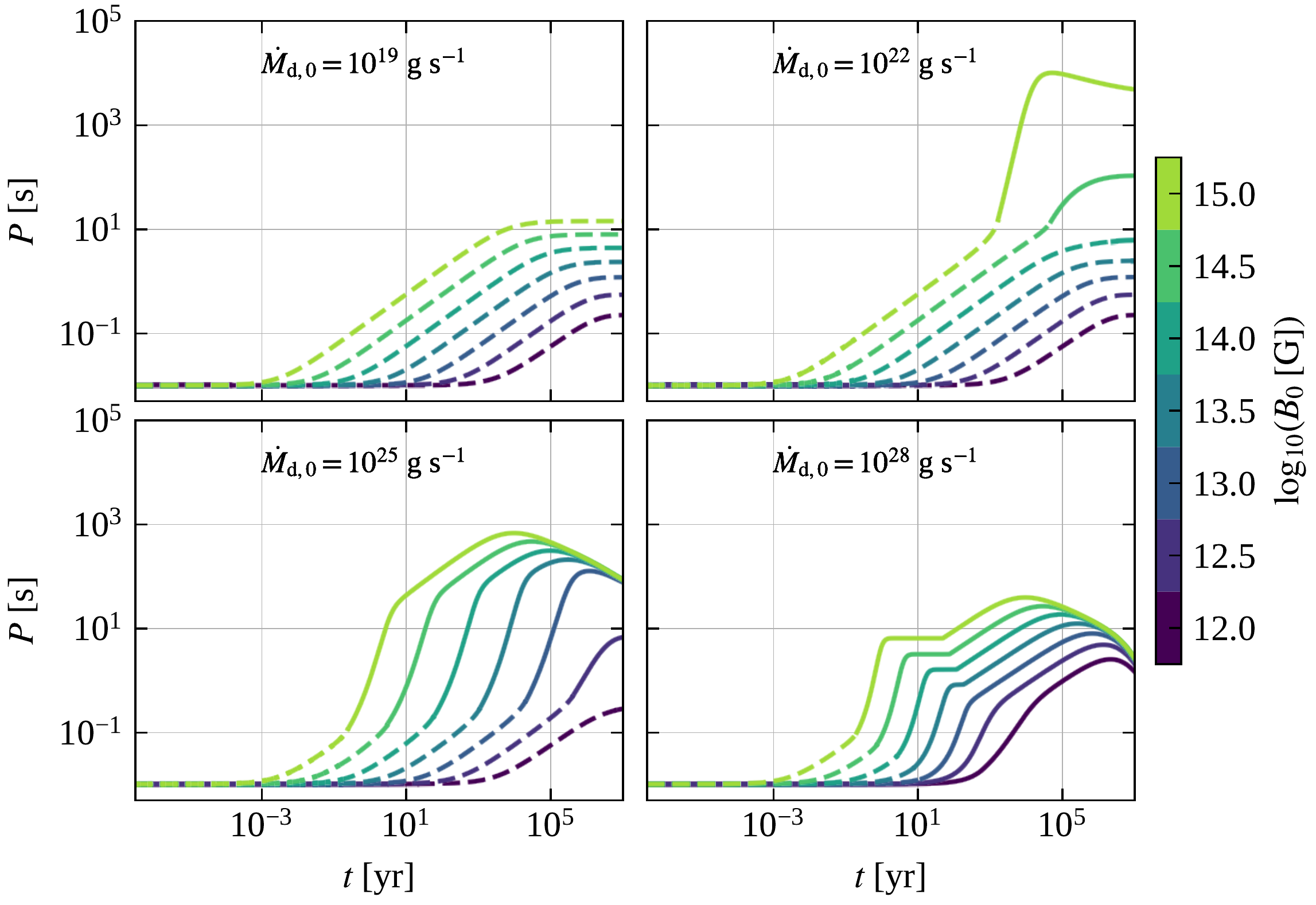}
\caption{Example curves showing the time evolution of the spin period for a pulsar interacting with a fallback disk on a timescale of $\unit[10^7]{yr}$, for different assumptions of the disk fallback rate $\dot{M}_{\rm d,0}$, and varying the initial magnetic field $B_0$. The dashed portion of the curves indicate when the neutron star is in the radio-loud ejector phase, i.e., when $r_{\rm m} > r_{\rm lc}$, while the solid portion indicates when the neutron star is in the radio-quiet propeller stage, i.e., when $r_{\rm m} < r_{\rm lc}$.}
\label{fig:P_evolution_Mdot}
\end{figure*}  

Figs.~\ref{fig:P_evolution_B} and \ref{fig:P_evolution_Mdot} show several examples of evolutionary curves of the spin period. We always assume an initial spin period $P_0 = \unit[10]{ms}$. In particular, Fig. \ref{fig:P_evolution_B} shows the spin-period evolution for several values of the initial magnetic field $B_0$ and varying initial disk accretion rate $\dot{M}_{\rm d,0}$, while Fig. \ref{fig:P_evolution_Mdot} shows the evolution curves for several values of initial disk accretion rate and varying initial magnetic field. In the early phases, the accretion at the inner radius of the disk is limited by the Eddington limit. 
For each simulated evolutionary curve we check that the disk's outer radius is always greater than the inner radius guaranteeing that the disk can form and influence the rotational evolution of the neutron star. In general it can be noted that for low values of the initial disk accretion rate ($< \unit[10^{22}]{g \, s^{-1}}$), the neutron star remains in the ejector phase for its entire evolution, independent of the value of its initial magnetic field. In contrast, higher accretion rates and higher magnetic fields allow the star to enter the propeller phase at earlier times. However, the propeller regime is most effective at spinning down the neutron star to periods $\gtrsim \unit[10]{s}$ only for relatively strong magnetic fields $\gtrsim \unit[10^{13}]{G}$ and intermediate initial disk accretion rates in the range $\unit[10^{22-27}]{g \, s^{-1}}$. 

We note that a neutron star that enters the propeller phase (and subsequently reaches spin equilibrium) will remain in this state until an abrupt change in the disk accretion rate occurs. For example, if the accretion rate in the disk suddenly drops, a neutron star can exit the propeller phase and enter the ejector phase again. In this case, the neutron star can transition from a faint X-ray source (due to thermal emission from material accreted onto the magnetosphere) to a standard rotation-powered radio pulsar or potentially radio-loud magnetar (see \S \ref{sec:discussion} for a detailed explanation of how this transition can occur and a description of the expected X-ray and radio luminosity in the different fall-back accretion states).

\subsection{Accretion from spherical fall-back}

If the fallback matter does not possess sufficient angular momentum to form a disk, fallback will proceed quasi-spherically. Even in this case, the neutron star could experience different accretion phases depending on the magnitude of the fallback rate. If the matter inflow is radiatively inefficient, accretion could proceed at super-Eddington rates, especially in the early stages. For such high fallback rates, the accretion flow is likely able to penetrate and squeeze the proto-neutron star magnetosphere, causing an initial phase of direct accretion onto the surface. This might also result in the burial of the magnetic field \citep[see for example][]{Taam1986, Li2021, Lin2021}, a scenario that (combined with the subsequent secular re-emergence of the magnetic field) has been invoked to explain the observed properties of Central Compact Objects (CCOs) \citep{Halpern2010, Fu2013, Ho2015, Zhong2021}; a class of young, generally weak-field neutron stars found close to the centers of supernova remnants. After this initial direct accretion phase, as the fallback rate decreases in time, the neutron star could enter a propeller phase that could cause the star to spin-down as in the disk scenario. However, in the case of quasi-spherical accretion, the fallback episode is expected to last at most a free-fall timescale $t_{\rm ff} \sim (G \rho)^{-1/2}$, determined by the density $\rho$ of the infalling outer layers of the progenitor participating in the fallback. In general, $t_{\rm ff}$ could reach at most $\sim \unit[10^7]{s}$ for red supergiant progenitor stars, whose envelopes have typical densities $\rho \sim \unit[10^{-7}]{g \, cm^{-3}}$ \citep{Metzger2018}.
Therefore, it is unlikely that a propeller phase acting on such short timescales (of the order of $\sim \unit[1]{yr}$) could result in equally long spin periods as the disk scenario.

\section{The 18 min periodic radio transient GLEAM-X\,J162759.5-523504.3}
\label{sec:gleam}

\begin{figure*}
\centering
\includegraphics[width=\textwidth]{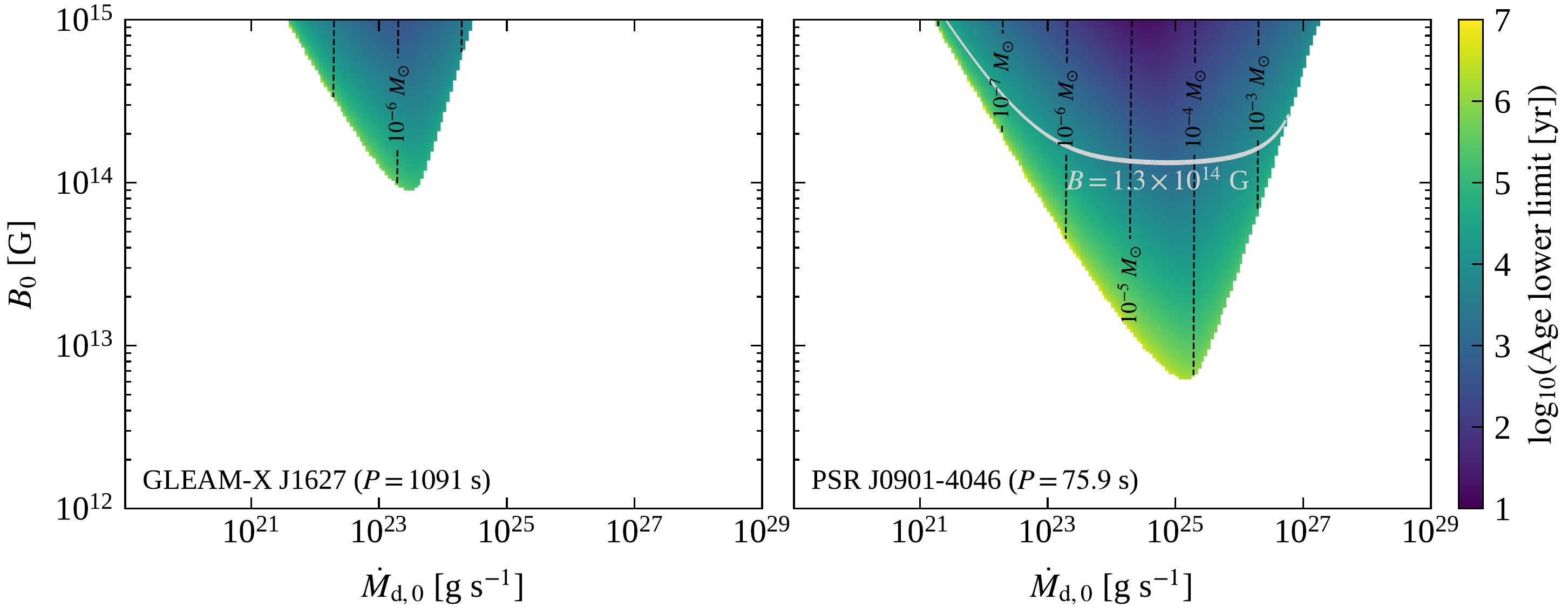}
\caption{Long-period pulsars in the supernova fallback-accretion picture. The colored regions show the values of the initial magnetic field $B_0$ and disk fallback rate $\dot{M}_{\rm d,0}$ that allow the neutron star to reach a spin period of $\unit[1091]{s}$ for \gleam\ (left panel) and of $\unit[75.9]{s}$ for \mtp\ (right panel) in less than $\unit[10^7]{yr}$. The color code indicates the time at which the neutron stars have reached their respective periods and could be interpreted as a lower limit on the source's current age. The contour lines indicate the total fallback mass that has been accreted by the disk in the same time interval. For \mtp, the gray contour line represents where the magnetic field value at $t=\rm Age$ is equal to the estimated magnetic field of \mtp\ from the spin-down formula.}
\label{fig:age_B_mdot}
\end{figure*}  

During  the GaLactic and Extragalactic All-sky MWA eXtended survey (GLEAM-X) \citep{Hurley-Walker2022b} with the Murchison Widefield Array (MWA), a peculiar periodic radio transient has been discovered displaying a periodicity of $\unit[1091]{s}$ \citep{Hurley-Walker2022a}. The source was detected during two radio outburst periods in January and March 2018, displaying a very variable flux (going from undetected to values as high as $\unit[20-50]{Jy}$), $\sim 5\%$ duty cycle, 90\% linear polarization, and a very spiky and variable pulse profile. 

From a detailed timing analysis, a dispersion measure of DM $=\unit[57\pm1]{pc \, cm^{-3}}$ was calculated, converting to a distance of $\unit[1.3]{kpc}$ according to the Galactic electron-density model of \citet{Yao2017}. The period derivative was loosely constrained to  $\dot{P}<\unit[1.5\times10^{-9}]{s \, s^{-1}}$. Assuming that the source is an isolated neutron star spinning down due to the classical electromagnetic dipole formula, this gives a relatively weak constraint on the dipolar magnetic field of $B_{\rm dip} <\unit[10^{17}]{G}$. 
We can also obtain an upper limit on the spin-down power $\dot{E}_{\rm rot} = I_{\rm NS} \omega \dot{\omega} < \unit[10^{28}]{erg \, s^{-1}}$. If we roughly estimate an isotropic radio luminosity from the flux detected during the outburst phase and the distance estimated from the DM, we obtain a value of $\sim \unit[10^{31}]{erg \, s^{-1}}$. From this, we deduce that the spin-down alone is insufficient to power these very bright radio flares.
The radio characteristics, such as the large flux variability, the radio outburst activity, and the high linear polarization, seem analogous to those observed in radio-active magnetars, whose activity is believed to be triggered and powered by the evolution and decay of their strong magnetic fields \citep{Thompson1995, Duncan1996}. However, \gleam's exceptionally long spin period would make this source stand out among them.

As shown in Fig. \ref{fig:em_spin_down_P_evol}, assuming \gleam\, is indeed a magnetar that has spun down via dipolar losses alone, in the limiting case of constant magnetic field, the magnetar would require an age $\gtrsim \unit[10^6 (10^8)] {yr}$ and a very strong magnetic field of $\unit[10^{16} (10^{15})]{G}$ to reach its spin period of $\unit[1091]{s}$. As argued in \S \ref{sec:dipolar_losses}, sustaining such a strong magnetic field over this lifetime is difficult to reconcile with crustal magnetic-field evolution models that predict field decay on the Hall timescale; in this case $\sim \unit[10^{3}]{yr}$. As a reference, \lowbsgr\ is most likely the oldest magnetar detected so far with a characteristic age of $\sim \unit[10^7]{yr}$ and has an inferred magnetic field of $\sim \unit[10^{13}]{G}$ \citep[Magnetar Outburst Online Catalogue \url{http:
//magnetars.ice.csic.es/}][]{CotiZelati2018}. Moreover, if the star's magnetic energy reservoir has decayed in time, it becomes challenging to explain the current magnetar-like activity observed from this source. 

A more promising explanation for \gleam\ that requires less extreme conditions is a magnetar that has experienced accretion from a fallback disk soon after the supernova. As outlined in \S \ref{sec:fallback_scenario}, a magnetar surrounded by a fallback disk will pass through the propeller phase and spin down very efficiently on short timescales. To study this scenario for \gleam, we first fix the initial spin period to $P_0 = \unit[10]{ms}$ (remember that as long as $P \gg P_0$, $P_0$ has very little influence on the long-term evolution); for the neutron star mass and radius we adopt the fiducial values $M_{\rm NS} = 1.4 \, M_{\odot}$ and $R_{\rm NS} = \unit[11]{km}$. By varying the two parameters $B_0$ and $\dot{M}_{\rm d,0}$ and using eq. \eqref{eq:torque} to determine the torque acting on the star in the different stages, we can numerically integrate eq. \eqref{eq:omega_evolution} in time. This allows us to find those parameter combinations that lead to a spin-down evolution reaching a period of at least $\unit[1091]{s}$. In what follows, we consider a maximum time of $\unit[10^7]{yr}$ for the evolution. The motivation for this limit is two-fold. Firstly, it ensures that the spin-period evolution curves reach their maximum values during the propeller phase before magnetic field decay enters into play (see the discussion of Figs.~\ref{fig:prop_evolution_ex}--\ref{fig:P_evolution_Mdot}). Secondly, a limit of $\unit[10^7]{yr}$ ensures that we encompass all age estimates of currently known magnetars (see the Magnetar Outburst Online Catalogue
\url{http://magnetars.ice.csic.es/} \citet{CotiZelati2018}).

The left panel in Fig. \ref{fig:age_B_mdot} shows the parameter space of $B_0$ and $\dot{M}_{\rm d,0}$, where we choose an initial magnetic field between $\unit[10^{12-15}]{G}$ and an initial disk accretion rate between $\unit[10^{19-29}]{g \, s^{-1}}$. The colored region represents the combination of parameters that guarantees the neutron star to reach a spin period of $\unit[1091]{s}$. The color code indicates the age at which the neutron star reaches the wanted period for the first time. The contour lines show the total fallback mass that has been accreted into the disk over the same time interval (calculated by integrating eq. \eqref{eq:disk_rate_evolution} in time). 
In this scenario, we observe that (thanks to the propeller phase) a magnetar with a magnetic field around $\unit[10^{14}]{G}$ can reach a spin period of $\unit[1091]{s}$ on a relatively short time scale of about $\unit[10^{3-5}]{yr}$ for an initial disk accretion rate of around $\unit[10^{23}]{g \, s^{-1}}$ and a total accreted mass of $\sim 10^{-6} M_{\odot}$. Note that since \gleam\ has been observed to emit periodic radio signals, disk accretion must have ceased to allow for the radio activity to be restored (see \S \ref{sec:discussion}). Assuming this is the case, the ages shown in Fig. \ref{fig:age_B_mdot} should be interpreted as the times at which the neutron star exited the propeller phase and shifted to standard dipolar spin-down; i.e., they represent lower limits on the real age of the observed source.

\section{The 76 seconds magnetar: \mtp}
\label{sec:mtp0013}

The long-period pulsar \mtp\, was recently discovered by the South African radio telescope MeerKAT. The pulsar manifests a periodicity of $P = \unit[75.9]{s}$ with a period derivative of $\dot{P} = \unit[2.25 \times 10^{-13}]{s \, s^{-1}}$ \citep{Caleb2022}. From the timing analysis a dispersion measure of ${\rm DM} = \unit[52 \pm 1]{pc\, cm^{-3}}$ has been derived which translates into a distance of $\sim \unit[0.4]{kpc}$.
Assuming that this neutron star is spinning down simply by dipolar losses, the strength of the dipolar magnetic field can be computed to $B_{\rm dip} \simeq \unit[1.3 \times 10^{14}]{G}$ and its spin-down power to $\dot{E}_{\rm rot} \simeq \unit[2 \times 10^{28}]{erg \, s^{-1}}$. Such a strong field is in the typical range of observed magnetars.

Using similar considerations as for \gleam\ and assuming that \mtp\ is an isolated neutron star, which has spun down due to electromagnetic dipolar losses alone, we can reproduce the observed period, if its magnetic field remained almost constant over a lifetime of almost $\unit[10^{7}]{yr}$ (the source's characteristic age is $\tau_c = P/(2 \dot{P}) \simeq \unit[5.4]{Myr}$). 
If we instead consider the fallback disk scenario, we can relax these conditions. 
As before, we solve the spin evolution equation eq. \eqref{eq:omega_evolution} for different combinations of the two parameters $B_0$ and $\dot{M}_{\rm d,0}$ and determine the parameter space that allows \mtp\ to reach a spin period of at least $P = \unit[75.9]{s}$ within $< \unit[10^{7}]{yr}$. The results are reported in the right panel of Fig. \ref{fig:age_B_mdot}. Also for \mtp, disk accretion must have stopped in order for the radio emission mechanism to be restored. From Fig. \ref{fig:age_B_mdot}, we deduce that observational characteristics can be explained within our model, if this happened at an age of around $\unit[10^{3-4}]{yr}$, which we again interpret as a lower limit on the pulsar's true age. This constraint on the age is in line with other radio emitting magnetars or high-$B$ radio pulsars that show similar radio emission. Furthermore in the case of \mtp, an initial disk accretion rate of around $\unit[10^{24}]{g \, s^{-1}}$, and a total accreted mass of $\sim 10^{-5} M_{\odot}$ are able to explain the observed period.

\section{Discussion} 
\label{sec:discussion}

\subsection{Re-establishing the radio emission after the propeller}
\label{subsec:disk_instability}

One key piece that we have to discuss is how a neutron star, that enters the propeller phase to experience effective spin-down, can exit this phase again in order to be observed as a radio-loud object; this is a crucial requirement, given that we need to explain the radio detection of \gleam\, and \mtp. As briefly mentioned in \S \ref{sec:fallback_scenario}, the transition from the propeller to the ejector phase can be naturally explained by an abrupt drop in the accretion rate. As a result, the magnetospheric radius would move beyond the light cylinder, providing the condition for the radio emission to be reactivated. 

Several factors could play a role in causing such a drop in the accretion flow. For example, since magnetars are very active young neutron stars that undergo outburst and flaring episodes \citep{CotiZelati2018, Esposito2021}, the energy released in such events could cause the disk to unbind or be completely disrupted. This would subsequently stop accretion and radio emission could be reinstated.

Another explanation could be simply that the fallback disk runs out of matter. During the propeller phase the material that reaches the magnetospheric radius is ejected due to the centrifugal barrier. If the ejected material possesses a velocity superior to the escape velocity, it will become unbound from the system; otherwise it will fall back and is reprocessed inside the accretion disk. As the matter feeding the disk from the supernova fallback is not replenished, if the propeller is efficient at unbinding matter from the system, we expect the accretion disk to eventually be completely consumed \citep[see for example][]{Eksi2005, Romanova2005}. 

Another possibility is that the fallback disk, which itself evolves with time, undergoes a thermal ionization instability, which has been outlined in detail in \citet{Mineshige1993, Menou2001, Ertan2009, Liu2015}. As the disk spreads and the accretion rate decays with time, energy dissipation decreases and the disk gradually cools down. In particular, \citet{Menou2001} argued that as the disk accretion rate falls below a critical value of around $\unit[10^{15}]{g \, s^{-1}}$ and the temperature in the outermost part of the disk drops below $\sim \unit[10^4]{K}$, the recombination of free electrons with heavy nuclei in the plasma is triggered. This transition alters the corresponding magnetic and viscous properties of the disk, reducing the efficiency of angular momentum transfer and eventually stopping accretion. As the disk evolves further in time, the recombination front propagates from the outermost regions inwards so that the active region of the disk shrinks. Eventually, if the disk becomes totally neutralized, accretion onto the neutron star will halt completely. \citet{Menou2001} have argued that this transition could occur around $\unit[10^{3-4}]{yr}$ when the outer disk radius is around $\sim \unit[10^{10-11}]{cm}$ 
This timescale is comparable to the ones we require in our fallback accretion scenario for neutron stars to reach spin periods $>\unit[10]{s}$ (see Fig. \ref{fig:age_B_mdot}). However, it has also been suggested that the irradiation from the central source may prevent the disk from becoming completely neutral, allowing it to stay active for longer times at even lower temperatures \citep[see for example][]{Alpar2001, Inutsuka2005, Alpar2013}. 
In general the evolution of a fallback disk is not trivial to model, since it also depends on the complex interaction with the central compact source. However, if the disk becomes inactive and the accretion flow stops due to any of the mechanisms outlined above, this could explain the transition from the propeller back to the ejector phase.

\begin{figure*}
\centering
\includegraphics[width = 0.9\textwidth]{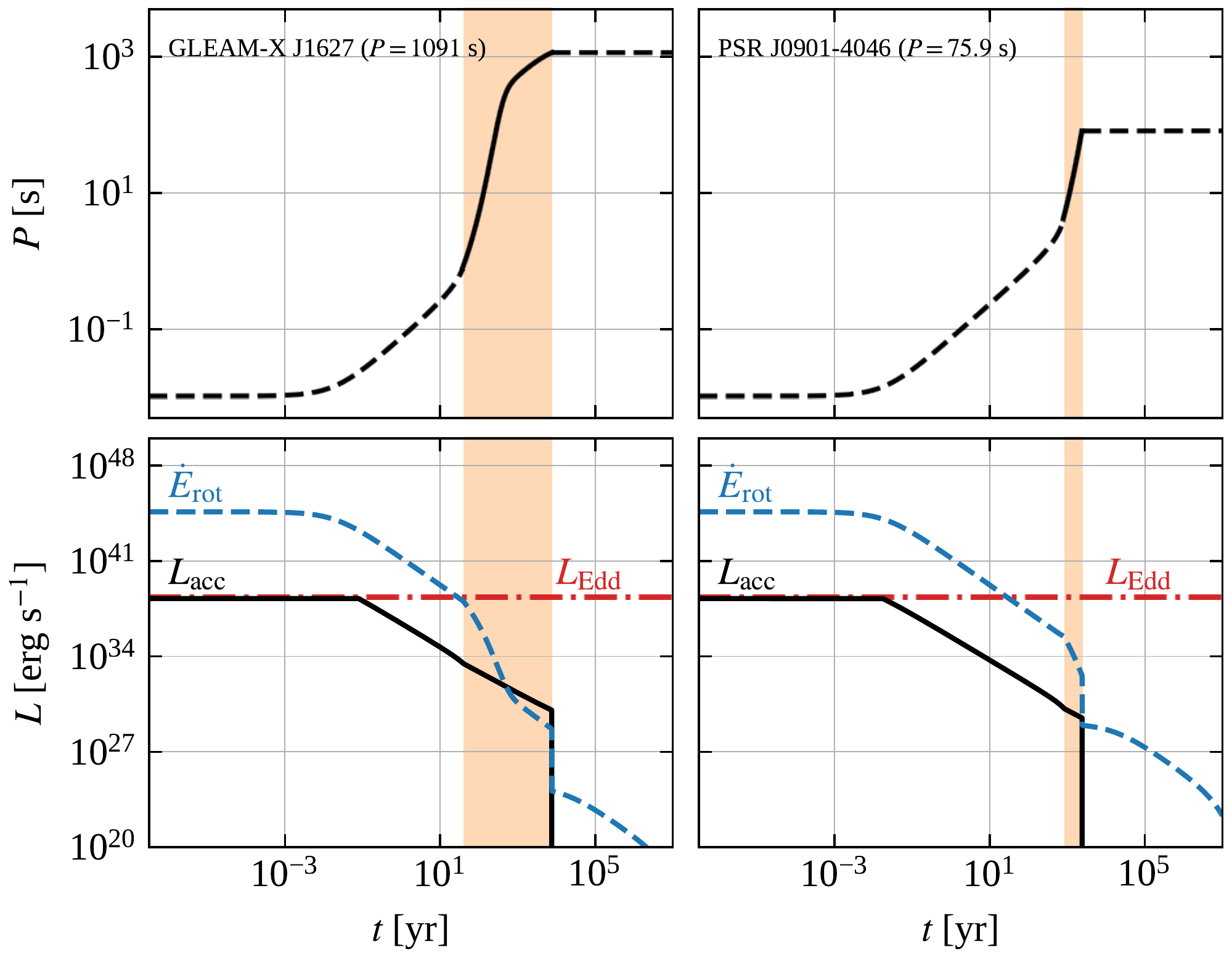}
\caption{Example of the time evolution of the spin period (top panels), disk luminosity and spin-down power (bottom panels) of \gleam\ (left panels) and \mtp\ (right panels). For both sources we assume an initial magnetic field $B_0=\unit[4 \times 10^{14}]{G}$ and an initial disk accretion rate of $\dot{M}_{\rm d,0} = \unit[10^{24}]{g \, s^{-1}}$ and $\dot{M}_{\rm d,0} = \unit[10^{23}]{g \, s^{-1}}$, respectively. The orange shaded region indicates the interval during which the two neutron stars spin down in the propeller phase. We assume that after the two neutron stars have reached their observed periods of $\unit[1091]{s}$ and $\unit[75.9]{s}$, respectively, an abrupt drop in accretion rate causes them to recover the ejector phase.}
\label{fig:GLEAM_MTP_evolution_L}
\end{figure*}  

In Fig. \ref{fig:GLEAM_MTP_evolution_L}, we show a possible evolutionary scenario for \gleam\ and \mtp\ that incorporates this drop in accretion rate. In particular, the top panels show the spin-period evolution. For both sources we choose an initial magnetic field $B_0=\unit[4 \times 10^{14}]{G}$ and initial disk accretion rates of $\dot{M}_{\rm d,0} = \unit[10^{24}]{g \, s^{-1}}$ and $\dot{M}_{\rm d,0} = \unit[10^{23}]{g \, s^{-1}}$, respectively. These values fall within the allowed parameter space that guarantees the two sources to reach their spin period in less than $\unit[10^7]{yr}$ (see Fig. \ref{fig:age_B_mdot}). Note that for \mtp, we fix the value of $B_0$ to a higher value than the estimated current magnetic field strength in order to take into account the effects of magnetic field decay.
To keep things simple, we set the disk accretion rate to zero as soon as the two neutron stars reach their observed spin periods of $\unit[1091]{s}$ and $\unit[75.9]{s}$, respectively. This abrupt stop in the accretion process causes the sources to transition into the ejector phase and enables the restoration of the radio emission. From this point on, the spin periods remain effectively constant as the combination of their large values and magnetic field decay causes the dipolar losses to become negligible.

On the other hand, if accretion would persist, the neutron stars would continue to evolve in a spin-equilibrium, unable to exit the propeller phase. This is possibly the evolutionary phase currently witnessed for \rcw, the $\unit[2]{kyr}$ old magnetar associated with the SNR RCW103. This source has shown magnetar-like outbursts \citep{Rea2016, D'Ai2016} and an observed period of $\unit[6.67]{hr}$ but is detected only in the X-ray band. Its period evolution has been modeled by \citet{Ho2017, Xu2019} using a similar model to our own. However, while \citet{Ho2017} assumed a constant accretion flow, in this work, we adopt the more realistic prescription of a time-varying accretion (similar to \citet{Xu2019}) as discussed in \S \ref{sec:fallback_scenario} 
In this scenario, the long period together with the young age of \rcw\ could be explained with a field of $\gtrsim \unit[10^{15}]{G}$ \citep[in line with the results of][]{Ho2017, Xu2019} and a relatively low initial disk accretion rate of $\sim \unit[10^{21}]{g  \, s^{-1}}$ (see bottom right panel of Fig. \ref{fig:P_evolution_B} as a reference). As \rcw\ is currently not observed in radio, and thus still in the propeller phase, we infer that the accretion disk is still active, which is compatible with the source's age of $\unit[2]{kyr}$.

An association of long-period pulsars with supernova remnants could present additional proof for the fallback disk scenario. However, for \gleam\ and \mtp\ no clear evidence of SNR associations has been discovered so far. In general, SNRs are expected to have an observational lifetime of around $\unit[10^{4-5}]{yr}$ \citep{Braun1989, Leahy2020}, which is comparable with our inferred timescales for the fallback disk to be active and accretion spinning down the two sources to their current periods. It is therefore possible that the associated SNRs are too faint to be detectable at the present time. Moreover, only around one third of the known young pulsars have a detected SNR remnant \citep[see the ATNF Pulsar Catalog][]{Manchester2005}. The reasons for this are still a matter of debate and study, but the absence of detected SNRs for most neutron stars could simply indicate differences in their progenitors, supernova explosions and interstellar environments \citep[see for example][]{Gaensler1995, Cui2021}. We therefore do not consider the lack of SNR associations for \gleam\ and \mtp\ as an issue for the validity of a fallback disk scenario. 

\subsection{Prediction for the X-ray and radio luminosity of long-period pulsars}
\label{subsec:luminosity}

In the bottom panels of Fig. \ref{fig:GLEAM_MTP_evolution_L}, we show the evolution of the disk luminosity and spin-down power for \gleam\ and \mtp. The accretion luminosity is computed as $L_{\rm acc} \simeq G M_{\rm NS} \dot{M}_{\rm d}/(2 r_{\rm in})$, where we recall that $r_{\rm in} \simeq \min(r_{\rm m}, r_{\rm lc})$. At very early times, i.e. $\lesssim \unit[1]{yr}$ after the supernova, the accretion flow is limited by the Eddington limit, suggesting that the disk emits in X-rays at the Eddington luminosity. As the system evolves and the disk accretion rate starts to decrease as $t^{-\alpha}$, the inner disk radius increases following the evolution of the magnetospheric radius as $r_{\rm in} \sim r_{\rm m} \propto t^{2 \alpha / 7}$. As a consequence, the luminosity decreases roughly $\propto t^{-9 \alpha/7}$. Once the disk becomes inactive or is completely consumed, the accretion rate is expected to vanish and the disk becomes undetectable in the X-rays.
Therefore assuming the fallback scenario for long-period radio transients such as \gleam\ and \mtp\ with a ceasing of the accretion flow, X-ray observations should not be able to detect emission from a residual disk if present. The cold debris of an inactive disk could instead be detectable in the infrared \citep{Wang2006, Posselt2018}. However, if the central neutron stars are indeed young (around $\unit[10^{5}]{yr}$) and have strong magnetic fields, they could emit thermal X-rays at their surfaces with luminosities up to $\sim \unit[10^{31-35}]{erg \, s^{-1}}$ due to the dissipation of magnetic energy in the crust \citep{Vigano2013}.

In Fig. \ref{fig:GLEAM_MTP_evolution_L} we also show the evolution of the spin-down power that is typically taken as the energy source for the radio emission of pulsars. In general, after these neutron stars have exited the propeller phase and recovered the ejector regime, we expect the spin-down to be caused only by electromagnetic torques so that $\dot{E}_{\rm rot} \propto B^2 / P^4$. Therefore, if we consider an upper limit for the magnetic field of around $\unit[10^{15}]{G}$ and a lower limit for the spin periods after the propeller regime of around $\unit[10]{s}$, for long-period pulsar we expect that the spin-down power has decayed to values $\lesssim \unit[10^{33}]{erg \, s^{-1}}$. This energy budget together with the magnetic energy stored in their strong fields could be enough to power radio emission and cause magnetar-like activity for neutron stars of this kind.

\section{Summary}
\label{summary}

We have studied the spin evolution of young, isolated neutron stars under the influence of fallback accretion. We specifically focused on the fallback-disk scenario as a promising origin of long-period pulsars, a class of objects that recent radio surveys are starting to unveil. 
By solving the torque balance equation for a disk accreting neutron star, we demonstrate that the evolution of such an object can differ significantly from standard dipole spin-down. In particular, we find that for a combination of high (but not extreme) magnetic field strengths and moderate fallback disk accretion rates in agreement with current core-collapse supernova simulations, neutron stars can enter the propeller phase during their evolution. This leads to effective spin-down and allows neutron stars to reach spin periods $\gg \unit[10]{s}$ on time scales on the order of $\sim \unit[10^{3-5}]{yr}$. Magnetic dipolar losses alone have problems to explain long spin periods and would require extreme conditions like strong and long-lasting magnetic fields potentially supported either by a core field component or other mechanisms such as the Hall attractor.

We have then interpreted the recently discovered objects \gleam\ and \mtp, with rotation periods of $\unit[1091]{s}$ and $\unit[75.9]{s}$, respectively, in light of this model, and showed that both objects could be explained as highly magnetized neutron stars with a fallback disk accretion history. The possibility to reach such long spin periods in much less than $\unit[10^{7}]{yr}$ is crucial to maintain the magnetic field and thus an energy reservoir to power their radio or X-ray activity. This is particularly important for \gleam, which was observed in outburst similar to other young radio-loud magnetars. 

We showed that for newly born neutron stars with birth fields of $B_0 \sim \unit[10^{14-15}]{G}$, a phase of fallback disk accretion with moderate initial accretion rates of $\unit[10^{22-27}]{g \, s^{-1}}$, could explain their detection as long-period radio or X-ray pulsars at relatively young ages ($\sim \unit[10^{3-5}]{yr}$). On the other hand, in systems where the initial magnetic fields are lower ($\sim \unit[10^{12-13}]{G}$), fallback disk accretion (even if present) is expected to have a negligible effect on the spin-period evolution. The majority of neutron stars will therefore primarily undergo standard dipolar electromagnetic spin-down and recover rotation periods below $\sim \unit[12]{s}$. In our framework, we therefore naturally recover the pulsar population which is observed to spin in the range $\sim$ 0.002--12~s. Note that the recently discovered radio pulsars PSR\,J1903$+$0433 \citep{Han2021} and PSR\,J0250$+$5854 \citep{Tan2018} with periods of $\unit[14]{s}$ and $\unit[23]{s}$, respectively, could be easily accommodated within our fallback accretion scenario. However, both sources can in principle also be explained within the standard evolutionary scenario provided that crustal field decay is very weak \citep[essentially requiring the absence of a highly resistive pasta layer; see][]{Pons2013}.

We also mentioned the X-ray emitting magnetar at the center of the $\unit[2]{kyr}$ old SNR RCW103, which requires (ongoing) fallback to explain its $\unit[6.67]{hr}$ period and radio-quiet nature. Finally note that classical magnetars with periods $\lesssim \unit[12]{s}$ would correspond to those systems where only strong fields are present, but fallback disk accretion does not take place or is inefficient because of a low $\dot{M}_{\rm d,0}$. 

In conclusion, fallback disk accretion after the supernova explosions of massive stars is expected to affect the evolution of newly born neutron stars. Depending on the relative intensities of the initial pulsar magnetic field and accretion rates, this scenario could represent an important ingredient to explain the connection between different neutron star classes and specifically shed light on the nature of the long-period radio sources recently discovered. 

\begin{acknowledgments}

We thank the anonymous referee for providing valuable comments that helped improve our manuscript. We are also grateful to Rosalba Perna for insightful discussions on fossil disk accretion and physics, and Daniele Vigan\'o for useful comments on this manuscript. We also thank Francesco Coti Zelati, Alice Borghese, Jos\'e Pons and Clara Dehman for useful discussions. We apologize to M. Caleb and her team for having unintentionally released sensible information on their work in an earlier unofficial release of this manuscript. MR's work has been carried out within the framework of the doctoral program in Physics of the Universitat Autònoma de Barcelona. MR, NR, and VG are supported by the H2020 ERC Consolidator Grant “MAGNESIA” under grant agreement No. 817661 (PI: Rea) and National Spanish grant PGC2018-095512-BI00. NHW is supported by an Australian Research Council Future Fellowship (project number FT190100231) funded by the Australian Government. 
This work was also partially supported by the program Unidad de Excelencia Mar\'ia de Maeztu CEX2020-001058-M, and by the PHAROS COST Action (No. CA16214).
\end{acknowledgments}


\%bibliography{bibliography}

\end{document}